\documentclass[showpacs,preprintnumbers,superscriptaddress]{revtex4}
\usepackage{CJK}
\usepackage{amsmath,amssymb,graphicx,bm}
\begin{document}
\begin{CJK*}{GBK}{song}

\title{Test the growth models of black hole by jointing LIGO and Insight-HXMT observations}

\author{Rong-Jia Yang \footnote{Corresponding author}}
\email{yangrongjia@tsinghua.org.cn}
\affiliation{College of Physical Science and Technology, Hebei University, Baoding 071002, China}
\affiliation{Hebei Key Lab of Optic-Electronic Information and Materials, Hebei University, Baoding 071002, China}
\affiliation{National-Local Joint Engineering Laboratory of New Energy Photoelectric Devices, Hebei University, Baoding 071002, China}
\affiliation{Key Laboratory of High-pricision Computation and Application of Quantum Field Theory of Hebei Province, Hebei University, Baoding 071002, China}

\author{Zhiwei Guo}
\email{yangrongjia@tsinghua.org.cn}
\affiliation{College of Physical Science and Technology, Hebei University, Baoding 071002, China}

\author{Yaoguang Zheng}
\affiliation{College of Physical Science and Technology, Hebei University, Baoding 071002, China}

\author{Shuang-Nan Zhang \footnote{zhangsn@ihep.ac.cn}}
\affiliation{Key Laboratory of Particle Astrophysics, Institute of High Energy Physics, Chinese Academy of Sciences, Beijing 100049, China}

\begin{abstract}
The growth model of black hole is still a controversial topic. In the stationary metric, all in-falling matter must be accumulated outside the event horizon of the black hole, as clocked by a distant external observer. In the time-dependent metric, all in-falling matter can fall into the event horizon of the final black hole within very short time. We test these two growth models by joining LIGO and Insight-HXMT observations. We find the stationary model is inconsistent with LIGO and Insight-HXMT observations.
\end{abstract}
\keywords{Gravitational Wave, Black Hole, GBR, Insight-HXMT}
\maketitle
\section{Introduction}
The detection of gravitational waves (GWs) by the LIGO Collaboration \cite{Abbott:2016blz} opens a new window for astronomical observations. GWs provide opportunities to test gravitational theories beyond general relativity. Up to now, no significant deviations from general relativity have been found in the weak-field regime. In a GW event, identifying the counterpart in electromagnetic (EM) band is also important to understand the full physical process. For example, we can probe extra dimension \cite{Yu:2016tar}, cosmological parameters \cite{Belgacem:2019tbw}, the weak equivalence principle \cite{Liu:2016edq}, and Lorentz invariance violations \cite{Abbott:2018lct} from joint gravitational wave and its counterpart observations.

Whether binary black holes (BBHs) merger produces electromagnetic counterpart is still an open problem. In general, generation of EM radiation in BBHs merger depends on the growth models of black hole (BH) and its environment. It has been argued that if BBHs merger happens in a gaseous disk around a supermassive BH, the process can be accompanied by a transient radio flare alike a fast radio burst \cite{Yi:2019rwo}. It was shown that EM signals can possibly follow from stellar-mass BH mergers \cite{deMink:2017msu}. GW events, short gamma-ray bursts, and fast radio bursts can also be expected from mergers of charged BHs \cite{Zhang:2016rli}. Generation of EM radiation in BBHs merger also depends on the growth models of BH. If one uses a stationary metric of a pre-existing BH to describe the growth of BH, one predicts strong EM radiation from merging BHs \cite{Vachaspati:2016hya, Zhang:2016kyq}. However, it argued in \cite{Liu:2009ts} that we should use time-dependent metric to describe the process for matters falling onto a BH, if we take into account the effect of in-falling matter on the background metric. This topic was further analysed in \cite{Zhao:2018ani,Zhang:2010vh}. In this model, all in-falling matters can fall into the BH within finite time. Therefore, one expects no EM radiation from merging BHs. These two contradictory conclusions are very interesting. Here we will test these two growth models of BH by jointing LIGO observations and Insight-HXMT data.

The rest of the paper is organised as follows. In section II, we briefly review the two growth models of BH. In section III, we test the growth models of BH by jointing LIGO observations and Insight-HXMT data. Finally, we briefly summarize our results in section V.

\section{Growth Models of black hole}
The growth of BH is a very important research topic, some aspects of it are still unclear. In BBHs merger, the production of EM radiation depends on the growth models of BH and its environment. In this section, we will briefly review two models about the growth of BH. Here we assume that the external observer (EO) is at rest and far away from the BH. From the viewpoint of an comoving observer (CO), the in-falling matter can cross the the event horizon (EH) of the BH. However, since the ``proper time" experienced by the in-falling CO can not be mapped back to the time of the external EO once the CO crossed the EH of the BH, therefore once the CO reached and then crossed the EH, whatever the CO experienced and ``saw" is irrelevant to the EO, so in all physical settings it is the viewpoint of EO that is relevant, for an observer on the earth. Second, since observed physical phenomena depends on the observer, when we compare two different physical models, for the sake of logical consistency, all our discussions can only be based on the same observer. Hence, as done in \cite{Vachaspati:2006ki,Liu:2009ts,Zhao:2018ani}, all our discussions, unless explicitly stated, refer to the coordinate time, which is the time experienced and recorded EO.

\subsection{Stationary model}
As first pointed out by Oppenheimer and Snyner, an EO sees the star asymptotically shrinking to its gravitational radius \cite{Oppenheimer:1939ue}, the collapsing body is forever suspended just above its Schwarzschild radius. That is to say, the in-falling matter will never reach the EH and certainly cannot cross the EH, seen from an EO. Therefore it must accumulate just outside the EH of the BH, as observed by the EO. Again emphasized in \cite{Vachaspati:2006ki}, it will take infinite time for the in-falling matter to reach the EH of a Schwarzschild BH, clocked by an EO. Since the metric is stationary in this theory, we call it as: stationary model. Taking into account the ``pre-Hawking radiation'' which can evaporate all in-falling matter in the collapsing process, so the BH will not grow \cite{Vachaspati:2006ki}. For any macro astrophysical object, however, the time scale of such evaporation process is much longer than the Hubble time \cite{Vachaspati:2006ki, Greenwood:2010sx}. Hence, its effects on the growth of any astrophysical BHs can be completely neglected. In the frame of stationary model, since all the in-falling matter is outside the EH of the BH, one expects strong EM radiation from merging BHs \cite{Vachaspati:2016hya, Zhang:2016kyq}.

\subsection{Time-dependent model}
There is another point of view about the growth of BH, as suggested in \cite{Liu:2009ts}.  For the whole gravitational system of an in-falling thick shell of matter onto a Schwarzschild BH, the metric must be time-dependent and the shell can cross the expanding EH within finite time, also clocked by a distant and rest EO \cite{Liu:2009ts}. Since the metric is time-dependent in this theory, we call this model as: time-dependent model. If there exist environmental materials, all of the in-falling matters can cross the final EH within finite time, seen from an EO. More specifically, the in-falling process can be divided into two stages. In the first stage, the matters fall via an accretion disk. But in the second stage, the matters are nearly free-fall after the innermost stable circular orbit. Therefore, the total time scale of the growth of BH is determined completely by the first stage \cite{Liu:2009ts}. In the frame of time-dependent model, one expects no EM radiation from merging BHs \cite{Liu:2009ts}.

\section{Tests by jointing LIGO observations and Insight-HXMT data}
In this section, we will test the two growth models of BH described above by jointing LIGO observations and Insight-HXMT data. The Hard X-ray Modulation Telescope (Insight-HXMT) is the first Chinese space X-ray telescope. It can monitor the entire GW localization area unblocked by the earth with microsecond time resolution in 0.2-5 MeV and very large collection area (~1000 cm$^{2}$) \cite{2020SCPMA..6349502Z}. Insight-HXMT can also quickly implement a Target of Opportunity observation to scan the GW localization area for potential X-ray emission from the GW source. So far, Insight-HXMT has made many interesting and important discoveries (see for example \cite{2021NatAs.tmp...54L, Wang:2021tcv, Ding:2021vuu, Ji:2020dnx, Kong:2020wvu}). Although Insight-HXMT did not detect any significant high energy (0.2-5 MeV) radiation from GW events, it provides most stringent constraints (~$10^{-7}$ to $10^{-6}$ erg/cm$^{2}$/s) for these GW events and any other possible precursor or extended emissions in 0.2-5 MeV \cite{2020SCPMA..6349502Z, Li:2017iup, Abbott:2016blz}.

\subsection{Stationary model}
In stationary model, since the matter accumulates outside their EHs, significant EM radiations are expected in the merging process of two BHs \cite{Vachaspati:2006ki, Vachaspati:2016hya}. However, the radiation will be gravitationally redshifted and diluted into unobservable level, so it will also take infinite time to observe such radiation for an EO \cite{Petrovay:2007qt}. If an accretion disk is formed, the angular momentum of the in-falling matter may compress and amplify magnetic fields around the EH of a BH during the in-falling process, which may produce jets and therefore significant EM radiation through a Blandford-Znajek (BZ)-like mechanism, avoiding the infinite redshift \cite{Blandford:1977ds}. Therefore, significant EM radiation may be observable even before the merging of the two BHs. If taking into account the spin of the BH and all the previously accumulated matter must be accreted onto it, more intensive EM radiation should be released during and after the merging process. Due to its high angular momentum, almost all of the accumulated matter could form a transient torus. The mass of this accretion disk can be estimated as follows. For the accretion disk, the magnetic pressure near the horizon may be limited by the inner disk pressure, so the magnetic field energy can be estimated by the disk pressure near the innermost stable circular orbit \cite{Beckwith:2009rm, Kawanaka:2012ub, Liu:2015lfa}. Taking into account the general relativistic effects and the contribution from the neutrino-optically thick region \cite{Gu:2006nu}, and including neutrino physics in the calculations, the magnetic pressure can be reckoned up through a numerical formulae found in \cite{Xue:2013boa, Liu:2015lfa}. Then, the BZ luminosity of a jet producing a weak short gamma-ray burst (GRB) can be approximated through a numerical formulae given in \cite{Lee:1999se}. This is the first way to obtain the BZ luminosity. Here we take the second way to calculate the BZ luminosity. Taking a common assumption that the magnetic field $B$ is around $10\%$ of its equipartition value \cite{Popham:1998ab, Matteo:2002ck, Mu:2018pky}, the analytic BZ jet power takes the form \cite{Popham:1998ab, Matteo:2002ck, Lee:1999md, Lee:1999se, Lei:2017zro, Mu:2018pky}
\begin{equation}
\label{e}
\dot{E}_{\mathrm{BZ}}=\lambda\left(a\right) \times 10^{51}\left(\frac{\dot{M}_{\mathrm{in}}}{M_{\odot}~ \mathrm{s}^{-1}}\right) \operatorname{erg} \mathrm{s}^{-1},
\end{equation}
where $\lambda$ is a function of $a$ which is the rotation parameter of the BH, and $\dot{M}_{\text {in }}=M_{\text {frag }}(1+z) / \omega$ is the average mass accretion rate. In \cite{Mu:2018pky}, some values for $\lambda(a)$ were found: $\lambda(a)=1.8$, $3.1$, and $4.4$ for $a = 0.8$, $0.9$, and $0.95$, respectively. The BZ jet power $\dot{E}_{\mathrm{BZ}}$ roughly equals the luminosity $L_{\rm BZ}$ of the flare \cite{Liu:2017rwh, Mu:2018pky}. Therefore from (\ref{e}), the BZ luminosity can be calculated with
\begin{equation}
\label{lu}
L_{\rm BZ}\simeq 4.4\times 10^{51}\dot{m}\,{\rm erg}\,{\rm s}^{-1}.
\end{equation}
where $\dot{m}=\dot{M}_{\mathrm{in}}/(M_{\odot}$ s$^{-1}$), and we take $\lambda(a)=4.4$ for $a=0.95$ obtained in \cite{Mu:2018pky}. The luminosity (\ref{lu}) can be collimated with an opening angle of $\theta_{\rm j}$ and is related to the observed isotropic radiation luminosity of the short GRB via
\begin{equation}
L_{\rm BZ}=\epsilon_\gamma^{-1}L_{\gamma,{\rm iso}}(1-\cos\theta_{\rm j}),
\end{equation}
where $L_{\gamma,{\rm iso}}$ is the observed isotropic radiation luminosity and $\epsilon_\gamma$ is the prompt radiation efficiency of the short GRB. We take the reported Insight-HXMT detection as the upper-limit $L_{\gamma, \rm{lim}}$ which are listed in Table \ref{tab1}. Then, we can derive the accretion rate
\begin{equation}
\dot{m}\leq \frac{L_{\gamma, \rm{lim}}}{4.4}\times 10^{-52}(\epsilon_\gamma/0.1)^{-1}(\theta_{\rm j}/0.1)^2,
\end{equation}
and the total mass of the accretion disk
\begin{equation}
M_{\rm torus}\leq \frac{L_{\gamma, \rm{lim}}}{4.4}\times 10^{-52} M_\odot(\epsilon_\gamma/0.1)^{-1}(\theta_{\rm j}/0.1)^2(\tau_\gamma/1\,{\rm s}),
\end{equation}
where $\tau_\gamma$ is the duration of the short GRB. Such a torus could be produced due to the fall-back of some ejecting materials during the supernova explosion of one pre-merging BH. Other possible ways by which this torus forms have been investigated in \cite{Perna:2016jqh,Loeb:2016fzn}.
\begin{table}[htbp]
  \centering
    \begin{tabular}{|c|c|c|c|c|c|c|c|}
    \hline
    LIGO/Virgo & Band model3   & Distance& $M_1$       & $M_2$       & Band model3  & $\dot{m}$ upper-limit &{$M_{\rm torus}$} \\
          & $\alpha$=-0.0         &(Mpc)    &($M_{\odot}$)&($M_{\odot}$)&upper-limit& accretion rate & upper-limit    \\
            &                     &        &              &             &Luminosity&    &($M_{\odot}$)    \\
          & $\beta$=-1.5          &         &             &             & (erg s$^{-1}$)&            &                \\
          & Ep=1000 keV        &         &             &             &              &                &                  \\
          &$3\sigma$ upper-limits&      &    &             &              &                          &              \\
          &fluence&      &    &             &              &                                          &       \\
          & (erg cm$^{-2}$) &         &             &             &              &                    &          \\
    ID    & 10s                &         &             &             &              &             &               \\
    \hline

    GW190408-181802 & 2.90$\times 10^{-7}$ & $1580^{+400}_{-590}$  & $24.5^{+5.1}_{-3.4}$ & $18.3^{+3.2}_{-3.5}$ & 2.12$^{+0.14}_{-0.30}\times 10^{49}$ & $4.82^{+0.32}_{-0.68}\times 10^{-4}$ &$4.82^{+0.32}_{-0.68}\times 10^{-3}$ \\
    \hline

    GW190412 & 4.60$\times 10^{-6}$ & $740^{+140}_{-170}$   & $30.0^{+4.7}_{-5.1}$ & $8.3^{+1.6}_{-0.9}$ & 7.21$^{+0.26}_{-0.38}\times 10^{49}$ & $1.64^{+0.58}_{-0.88}\times 10^{-3}$& $1.64^{+0.58}_{-0.88}\times 10^{-2}$ \\
    \hline

    GW190512-180714 & 1.90$\times 10^{-6}$ & $1490^{+530}_{-590}$  & $23.0^{+5.4}_{-5.7}$ & $12.5^{+3.5}_{-2.5}$ & 1.20$^{+0.15}_{-0.19}\times 10^{50}$ & $2.7^{+0.37}_{-0.4}\times 10^{-3}$ &$2.7^{+0.37}_{-0.4}\times 10^{-2}$\\
    \hline

    GW190513-205428 & 2.80$\times 10^{-6}$ & $2160^{+940}_{-800}$  & $35.3^{+9.6}_{-9.0}$ & $18.1^{+7.3}_{-4.2}$ & 2.96$^{+0.56}_{-0.41}\times 10^{50}$ & $6.73^{+1.27}_{-0.93}\times 10^{-3}$ &$6.73^{+1.27}_{-0.93}\times 10^{-2}$ \\
    \hline

    S190517-055101 & 3.40$\times 10^{-6}$ & $2110^{+1790}_{-1000}$  & $36.4^{+11.8}_{-7.8}$ & $24.8^{+6.9}_{-7.1}$ & 2.08$^{+1.50}_{-0.47}\times 10^{50}$ & $4.73^{+3.41}_{-1.07}\times 10^{-3}$ &$4.73^{+3.41}_{-1.07}\times 10^{-2}$ \\
    \hline

    GW190519-153544 & 1.40$\times 10^{-6}$ & $2850^{+2020}_{-1140}$  & $64.5^{+11.3}_{-13.2}$ & $39.9^{+11.0}_{-10.6}$ & 3.11$^{+1.56}_{-0.50}\times 10^{50}$ & $7.07^{+3.55}_{-1.13}\times 10^{-3}$ &$7.07^{+3.55}_{-1.13}\times 10^{-2}$ \\
    \hline

    GW190521 & 1.40$\times 10^{-6}$ & $4530^{+2300}_{-2130}$  & $91.4^{+29.3}_{-17.5}$ & $66.8^{+20.7}_{-20.7}$ & 7.86$^{+2.03}_{-1.74}\times 10^{50}$ & $1.79^{+0.46}_{-0.39}\times 10^{-2}$ &$1.79^{+0.46}_{-0.39}\times 10^{-1}$ \\
    \hline

    GW190602-175927 & 1.30$\times 10^{-6}$ & $2990^{+2020}_{-1260}$ & $67.2^{+16.0}_{-12.6}$ & $47.4^{+13.4}_{-16.6}$ & 3.85$^{+1.76}_{-0.68}\times 10^{50}$ &$8.75^{+4.0}_{-3.87}\times 10^{-3}$ & $8.75^{+4.0}_{-3.87}\times 10^{-2}$ \\
    \hline

    GW190630-185205 & 4.50$\times 10^{-6}$ & $ 930^{+560}_{-400}$ & $35.0^{+6.9}_{-5.7}$ & $23.6^{+5.2}_{-5.1}$ & 5.48$^{+1.99}_{-1.01}\times 10^{49}$ & $1.25^{+0.45}_{-0.23}\times 10^{-3}$ &$1.25^{+0.45}_{-0.23}\times 10^{-2}$ \\
    \hline

    GW190701-203306 & 7.20$\times 10^{-7}$ & $2140^{+790}_{-730}$  & $53.6^{+11.7}_{-7.8}$ & $40.8^{+8.3}_{-11.5}$ & 9.31$^{+1.27}_{-1.08}\times 10^{49}$ & $2.12^{+0.28}_{-0.25}\times 10^{-3}$ &$2.12^{+0.28}_{-0.25}\times 10^{-2}$ \\
    \hline

    GW190706-222641 & 6.80$\times 10^{-7}$ & $5070^{+2570}_{-2110}$  & $64.0^{+15.2}_{-15.2}$ & $38.5^{+12.5}_{-12.4}$ & 7.07$^{+1.82}_{-1.23}\times 10^{50}$ & $1.61^{+0.41}_{-0.28}\times 10^{-2}$ & $1.61^{+0.41}_{-0.28}\times 10^{-1}$\\
    \hline

    GW190707-093326 & 1.40$\times 10^{-6}$ & $800^{+370}_{-380}$   & $11.5^{+3.3}_{-1.7}$ & $8.4^{+1.4}_{-1.6}$ & 2.14$^{+0.46}_{-0.48}\times 10^{49}$ & $4.86^{+1.05}_{-1.09}\times 10^{-4}$ &$4.86^{+1.05}_{-1.09}\times 10^{-3}$ \\
    \hline

    GW190828-065509 & 2.80$\times 10^{-6}$ & $1660^{+630}_{-610}$  & $23.8^{+7.2}_{-7.0}$ & $10.2^{+3.5}_{-2.1}$ & 2.90$^{+0.42}_{-0.39}\times 10^{50}$ & $6.59^{+0.96}_{-0.89}\times 10^{-3}$ &$6.59^{+0.96}_{-0.89}\times 10^{-2}$ \\
    \hline

    GW190930-133541 & 1.30$\times 10^{-6}$ & $780^{+370}_{-330} $  & $12.3^{+12.5}_{-2.3}$ & $7.8^{+1.7}_{-3.3}$ & 3.35$^{+0.75}_{-0.60}\times 10^{49}$ & $7.61^{+1.71}_{-1.36}\times 10^{-4}$ &$7.61^{+1.71}_{-1.36}\times 10^{-3}$ \\
    \hline
    \end{tabular}
     \caption{GW events and Insight-HXMT constraints: $3\sigma$ upper-limits fluence constrained from Insight-HXMT, the distances of sources, the masses of BBHs, the upper-limit luminosity, and the $M_{\rm torus}$ upper-limit.}
\label{tab1}
\end{table}%
Assuming the GW counterpart GRB with three typical GRB Band spectral models and two typical duration timescales (1 s, 10 s) from the center of the LIGO-Virgo location probability map, Insight-HXMT provides 3$\sigma$ upper-limits fluence (0.2-5 MeV, incident energy) for each Band at each duration timescales. Insight-HXMT, for example, has provided one of the most stringent constraints (~$10^{-7}$ to $10^{-6}$ erg/cm$^2$/s) for both GRB170817A and any other possible precursor or extended emissions in 0.2-5 MeV \cite{Li:2017iup}. In these Band models, we find the values of upper-limits fluence for Band 3 at duration timescales 10 s are the highest, shown in Table \ref{tab1}. So we use these data to calculate the upper-limits luminosity of GRB and therefore the total upper-limits mass of the accretion disk.

The estimated mass $M_{\rm torus}$ upper-limits accumulated around the BH in each GW event are listed in table \ref{tab1}; they are all less than one solar mass, significantly smaller than the mass of any one of the corresponding pre-merging BH in the related GW event: $M_{\rm torus}\ll {\rm min} \{M_1, M_2\}$. Compared with the mass of the corresponding pre-merging BH, the estimated mass $M_{\rm torus}$ is almost negligible. In these 14 GW events, the mass of any one of the pre-merging BH is several or tens of solar masses, while a stellar mass BH is commonly believed to be formed as a result of stellar evolution, as evidenced by the mass distributions of the stellar mass BH in X-ray binaries \cite{Zhang:2013fwa}. However, the masses of the two BHs found in GW190521 are much larger than the masses of stellar mass BHs in all known X-ray binaries. Therefore, they should have significantly grown by accretion after their births, e.g., from the companion star, the fall-back gas, or the interstellar medium. It has been shown, for example, that three distinct possible types of the GW190521 original progenitors formed at lower masses and grew to their estimated LVC parameters by relativistic accretion ware found: (i) $10^{-4}M_{\odot}-3M_{\odot}$ primordial BHs, for an early universe, $z\sim 100$, origin; (ii) $3M_{\odot}-40M_{\odot}$ stellar mass BHs at $z\sim 50$; (iii) $40M_{\odot}-60M_{\odot}$ BHs at $z\sim 20$, which could originate from the collapse of high mass Pop III stars \cite{Cruz-Osorio:2021qbr}. The formation of binaries in GW190521 is also possible through gas accretion onto the BH remnants of Population III stars born in high-redshift ($z > 10$) minihalos \cite{Safarzadeh:2020vbv}. In dense gas-rich nuclear star clusters, wind-fed supraexponential accretion under the assumption of net zero angular momentum for the gas, can lead to extremely rapid growth, scaling stellar mass remnant seed BHs up to the intermediate mass BH range \cite{Natarajan2009}. In \cite{Woosley:2021xba}, four factors affecting the theoretical estimates for the boundaries of GW190521's mass gap were explored: uncertainties in reaction rates by themselves allow the mass below some limiting value, $M_{\rm{low}}$, rise to 64 $M_{\odot}$; rapid rotation could increase $M_{\rm{low}}$ to $\sim 70 M_{\odot}$; super-Eddington accretion and evolution in detached binaries can increase $M_{\rm{low}}$ still further. In these scenarios, the matters accumulated in the stationary model by the two pre-merging BHs in GW190521 is far greater than the mass of torus $M_{\rm torus}$ estimated from LIGO and Insight-HXMT data. Therefore, we conclude that the stationary model is not reasonable.

\subsection{Time-dependent model}
Now we consider the time-dependent model. In the 14 GW events above detected by the LIGO Scientific Collaboration \& the Virgo Collaboration, the mass of pre-merging BH is in the range of several to tens of solar masses. As evidenced by the mass distributions of the stellar mass black holes in X-ray binaries \cite{Zhang:2013fwa}, a stellar mass black hole is commonly believed to be formed as a result of stellar evolution. Despite of different accretion modes, it will take about $\sim10^9$ yr for a BH with several solar mass in an X-ray binary to grow to tens solar mass under the typical accretion rate ($\sim10^{-8} M_\odot$) of an X-ray binary. It is possible, however, that the BBHs may have grown up in much denser environment with much higher accretion rate, $\sim10^9$ yr can be considered a rough upper limit for the time scale of the growth of BH. If the decay of the orbit driven by the energy and angular momentum loss because of the emission of gravitational waves, the time to merger for BBHs with component masses $M_1$, $M_2$, and separation $d$ is given \cite{Peters:1964zz}
\begin{equation}
\tau_\textrm{GW} = \frac{5}{256} \frac{c^5}{G^3} \frac{d^4}{M_1 M_2 (M_1+M_2)}\, .
\end{equation}
It has been argued that the coalescence time needed for two solar masses BHs, like the GW150914 and GW190521 events, is typically in the range of several to 11 Gyr after formation \cite{Mandel:2015qlu, Belczynski:2016obo, Natarajan2009, Cruz-Osorio:2021qbr, Safarzadeh:2020vbv}. This means, long before the final merging stage of the solar masse BHs in the GW150914 and GW190521 events, practically most of the matter should have disappeared into the two BHs. This was explicitly predicted in time-dependent model \cite{Liu:2009ts}. This conclusion is also generic and independent of where the in-falling matter comes from and how the BHs were formed, because in time-dependent model the in-falling matter will disappear into the expanding EH within time scales much shorter than any other time scales involved in producing the final merging event, as far as they were formed through astrophysical collapse \cite{Liu:2009ts}.

As a matter of fact, most of the known stellar mass BHs in X-ray binaries are transients, i.e., they spend most of their lives in quiescence with mass accretion rate much lower than $\sim10^{-8} M_\odot$ and only occasionally show bright outbursts with mass accretion rate on the order of $\sim10^{-8} M_\odot$. This would imply that the growth time scale of the BHs of these GW events might be much longer than $\sim10^9$ yr, if these BHs are formed this way. However, this still does not change our conclusion above, since the low mass accretion rate ensures that negligible amount of matter is left outside the event horizon right before the final merger, as the previously accreted matter has continuously disappeared into the expanding event horizon. On the other hand, rapid BH growth seems preferred in essentially all models for growing these BHs, as we have discussed in the end of the previous subsection (Section III A). In this scenario, there is also nothing left outside the BH's event horizon right before the final merger, in the framework of this time dependent model. Therefore our conclusion remains valid for all known scenarios of BH growth.

\section{Conclusion and discussion}
We investigated two possible growth models of the stellar mass BHs in 14 GW events detected by LIGO, as clocked by an EO. In stationary model, one describes the growing process of BH through matter falling into a seed BH by using a stationary metric, so all in-falling matter must be accumulated outside the EH of the BH, within finite time clocked by the EO. Taking the Insight-HXMT reported GRB luminosity as upper limits to the possible GRB associated with the GW events, we found that the mass accumulated outside the EH of the BH is significantly smaller than that accumulated by the corresponding pre-merging BH in each GW event, implying that the stationary model is inconsistent with LIGO and Insight-HXMT observations.

In time-dependent model, practically all in-falling matter can cross the EH of the BH within short time, also clocked by an EO. If an accretion disk around the BH is formed, the total accretion time scale is still much shorter than the coalescence time needed for the two BHs in the GW190521 event for any reasonable binary orbital radius, provided the only orbital energy loss mechanism is through releasing GW energy. Therefore, long before the final merging stage of the two BHs like in the GW190521 event, practically all matter should have fallen into the two BHs. So in this scenario, it's less likely that EM radiations can be expected from the BBHs merging, which was explicitly predicted in \cite{Liu:2009ts}.

\begin{acknowledgments}
We thank Y. Liu for helpful discussions. This work is supported in part by Hebei Provincial Natural Science Foundation of China (Grant No. A2014201068).
\end{acknowledgments}

\bibliographystyle{ieeetr}
\bibliography{ref}
\end{CJK*}
\end{document}